\documentclass{article}
\title{Perturbative Gadgets at Arbitrary Orders}
\author{Stephen P. Jordan\footnote{stephen.jordan@nist.gov} \ and Edward
  Farhi\footnote{farhi@mit.edu} \vspace{5pt} \\
\normalsize{\it{MIT Center for Theoretical Physics}}}
\date{}

\usepackage{graphicx}
\usepackage{amssymb}
\usepackage{amsmath}
\usepackage{bm}
\usepackage{fullpage}
\usepackage{dsfont}

\newcommand{\captionfonts}{\small}

\makeatletter  
\long\def\@makecaption#1#2{%
  \vskip\abovecaptionskip
  \sbox\@tempboxa{{\captionfonts #1: #2}}%
  \ifdim \wd\@tempboxa >\hsize
    {\captionfonts #1: #2\par}
  \else
    \hbox to\hsize{\hfil\box\@tempboxa\hfil}%
  \fi
  \vskip\belowcaptionskip}
\makeatother   

\begin{document}
\bibliographystyle{plain}
\maketitle
\newcommand{\ud}{\mathrm{d}}
\newcommand{\bra}[1]{\langle #1|}
\newcommand{\ket}[1]{|#1\rangle}
\newcommand{\braket}[2]{\langle #1|#2\rangle}
\newcommand{\Bra}[1]{\left<#1\right|}
\newcommand{\Ket}[1]{\left|#1\right>}
\newcommand{\Braket}[2]{\left< #1 \right| #2 \right>}
\renewcommand{\th}{^\mathrm{th}}
\newcommand{\tr}{\mathrm{Tr}}
\newcommand{\cu}{\mathcal{U}}
\newcommand{\ca}{\mathcal{A}}
\newcommand{\id}{\mathds{1}}
\newcommand{\hv}{\hat{V}}

\newtheorem{lemma}{Lemma}
\newtheorem{theorem}{Theorem}
\newtheorem{prop}{Proposition}

\begin{abstract}
Adiabatic quantum algorithms are often most easily formulated using
many-body interactions. However, experimentally available interactions
are generally two-body. In 2004, Kempe, Kitaev, and Regev introduced
perturbative gadgets, by which arbitrary three-body effective
interactions can be obtained using Hamiltonians consisting only of
two-body interactions. These three-body effective interactions arise
from the third order in perturbation theory. Since their introduction,
perturbative gadgets have become a standard tool in the theory of
quantum computation. Here we construct generalized gadgets so that one
can directly obtain arbitrary $k$-body effective interactions from two-body
Hamiltonians. These effective interactions arise from the $k\th$ order
in perturbation theory.
\end{abstract}

\section{Perturbative Gadgets}
\label{intro}

Perturbative gadgets were introduced to construct a two-local Hamiltonian
whose low energy effective Hamiltonian corresponds to a desired
three-local Hamiltonian. They were originally
developed by Kempe, Kitaev, and Regev in 2004 to prove the
QMA-completeness of the 2-local Hamiltonian problem and to simulate 
3-local adiabatic quantum computation using 2-local adiabatic quantum
computation~\cite{Kempe}. Perturbative gadgets have subsequently been
used to simulate spatially nonlocal Hamiltonians using spatially local
Hamiltonians~\cite{Oliveira}, and to find a minimal set of set of
interactions for universal adiabatic quantum
computation~\cite{Biamonte}. It was also pointed out in~\cite{Oliveira}
that perturbative gadgets can be used recursively to obtain $k$-local
effective interactions using a 2-local Hamiltonian. Here we generalize
perturbative gadgets to directly obtain arbitrary $k$-local effective
interactions by a single application of $k\th$ order perturbation
theory. Our formulation is based on a perturbation expansion due to
Bloch~\cite{Bloch}. 

A $k$-local operator is one consisting of interactions between at most
$k$ qubits. A general $k$-local Hamiltonian on $n$ qubits can always
be expressed as a sum of $r$ terms,
\begin{equation}
\label{hcomp}
H^{\mathrm{comp}} = \sum_{s=1}^r c_s H_s
\end{equation}
with coefficients $c_s$, where each term $H_s$ is a $k$-fold
tensor product of Pauli operators. That is, $H_s$ couples some set of
$k$ qubits according to
\begin{equation}
H_s = \sigma_{s,1} \ \sigma_{s,2} \ \ldots \ \sigma_{s,k},
\end{equation}
where each operator $\sigma_{s,j}$ is of the form
\begin{equation}
\sigma_{s,j} = \hat{n}_{s,j} \cdot \vec{\sigma}_{s,j},
\end{equation} 
where $\hat{n}_{s,j}$ is a unit vector in $\mathbb{R}^3$, and 
$\vec{\sigma}_{s,j}$ is the vector of Pauli matrices operating on the
$j\th$ qubit in the set of $k$ qubits acted upon by $H_s$.

We wish to simulate $H^{\mathrm{comp}}$ using only 2-local
interactions. To this end, for each term $H_s$, we introduce $k$
ancilla qubits, generalizing the technique of~\cite{Kempe}. There are
then $rk$ ancilla qubits and $n$ computational qubits. We choose
the gadget Hamiltonian as
\begin{equation}
H^{\mathrm{gad}} = \sum_{s=1}^r H_s^{\mathrm{anc}} + \lambda
\sum_{s=1}^r V_s,
\end{equation}
where
\begin{equation}
\label{nonpert}
H_s^{\mathrm{anc}} = \sum_{1 \leq i < j \leq k } \frac{1}{2} 
(I - Z_{s,i} Z_{s,j}),
\end{equation}
and
\begin{equation}
\label{pert}
V_s = \sum_{j = 1}^k c_{s,j} \sigma_{s,j} \otimes X_{s,j}
\end{equation}
and
\begin{equation}
c_{s,j} = \left\{ \begin{array}{ll}
c_s & \textrm{if $j=1$} \\
1 & \textrm{otherwise.}
\end{array} \right.
\end{equation}
For each $s$ there is a corresponding register of $k$ ancilla
qubits. The operators $X_{s,j}$ and $Z_{s,j}$ are Pauli $X$ and
$Z$ operators acting on the $j\th$ ancilla qubit in the ancilla
register associated with $s$. For each ancilla register, the ground
space of $H_s^{\mathrm{anc}}$ is the span of $\ket{000\ldots}$ and
$\ket{111\ldots}$. $\lambda$ is the small parameter in which the
perturbative analysis is carried out.

\begin{figure}
\begin{center}
\includegraphics[width=0.15\textwidth]{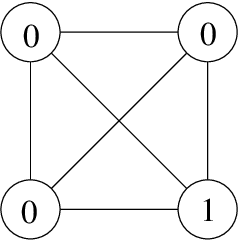}
\caption{\label{ancillas} The ancilla qubits are all coupled together
  using $ZZ$ couplings. This gives a unit energy penalty for each pair
  of unaligned qubits. If there are $k$ bits, of which $j$ are in the
  state $\ket{1}$ and the remaining $k-j$ are in the state $\ket{0}$,
  then the energy penalty is $j(k-j)$. In the example shown in this
  diagram, the 1 and 0 labels indicate that the qubits are in the
  state $\ket{0001}$, which has energy penalty 3.}
\end{center}
\end{figure}

For each $s$, the operator 
\begin{equation}
X_s^{\otimes k} = X_{s,1} \otimes X_{s,2} \otimes \ldots \otimes X_{s,k}
\end{equation}
acting on the $k$ ancilla qubits in the 
register $s$ commutes with $H^{\mathrm{gad}}$. Since there are $r$
ancilla registers, $H^{\mathrm{gad}}$ can be block diagonalized into
$2^r$ blocks, where each register is in either the $+1$ or $-1$
eigenspace of its $X_s^{\otimes k}$. In this paper, we analyze only the 
block corresponding to the $+1$ eigenspace for every register. This
$+1$ block of the gadget Hamiltonian is a Hermitian operator,
that we label $H_+^{\mathrm{gad}}$. We show that the
effective Hamiltonian on the low energy eigenstates of
$H_+^{\mathrm{gad}}$ approximates $H^{\mathrm{comp}}$. For many
purposes this is sufficient. For example, suppose one wishes to
simulate a $k$-local adiabatic quantum computer using a $2$-local
adiabatic quantum computer. If the initial state of the computer lies
within the all $+1$ subspace, then the system will remain in this
subspace throughout its evolution. To put the initial state of the
system into the all $+1$ subspace, one can initialize each ancilla
register to the state
\begin{equation}
\ket{+} = \frac{1}{\sqrt{2}} (\ket{000\ldots} + \ket{111\ldots}),
\end{equation}
which is the ground state of $\sum_s H_s^{\mathrm{anc}}$ within the
$+1$ subspace. Given the extensive experimental literature on the
preparation of states of the form $\ket{+}$, sometimes called cat states,
a supply of such states seems a reasonable resource to assume.

The purpose of the perturbative gadgets is to obtain $k$-local
effective interactions in the low energy subspace. To quantify this,
we use the concept of an effective Hamiltonian. We define this
to be
\begin{equation}
H_{\mathrm{eff}}(H,d) \equiv \sum_{j=1}^d E_j \ket{\psi_j}\bra{\psi_j},
\end{equation}
where $\ket{\psi_1},\ldots,\ket{\psi_d}$ are the $d$ lowest energy
eigenstates of a Hamiltonian $H$, and $E_1, \ldots , E_d$ are
their energies.

In section \ref{section_analysis}, we calculate
$H_{\mathrm{eff}}(H^{\mathrm{gad}}_+,2^n)$ perturbatively to $k\th$
order in $\lambda$. To do this, we write $H^{\mathrm{gad}}$ as
\begin{equation}
H^{\mathrm{gad}} = H^{\mathrm{anc}} + \lambda V
\end{equation}
where
\begin{equation}
H^{\mathrm{anc}} = \sum_{s=1}^r H_s^{\mathrm{anc}}
\end{equation}
and
\begin{equation}
V = \sum_{s=1}^r V_s
\end{equation}
We consider $H^{\mathrm{anc}}$ to be the unperturbed Hamiltonian and
$\lambda V$ to be the the perturbation. We find that $\lambda V$
perturbs the ground space of $H^{\mathrm{anc}}$ in two separate
ways. The first is to shift the energy of the entire space. The second
is to split the degeneracy of the ground space. This splitting arises
at $k\th$ order in perturbation theory, because the lowest power of
$\lambda V$ that has nonzero matrix elements within the ground space
of $H^{\mathrm{anc}}$ is the $k\th$ power. It is this splitting which
allows the low energy subspace of $H^{\mathrm{gad}}_+$ to mimic the
spectrum of $H^{\mathrm{comp}}$.

It is convenient to analyze the shift and the splitting
separately. To do this, we define 
\begin{equation}
\widetilde{H}_{\mathrm{eff}}(H,d,\Delta) \equiv H_{\mathrm{eff}}(H,d)
- \Delta \Pi,
\end{equation}
where $\Pi$ is the projector onto the subspace
\begin{equation}
\label{E-quation}
\mathcal{E} = \mathrm{span}\{\ket{\psi_1},\ldots,\ket{\psi_d}\}.
\end{equation}
Thus, $\widetilde{H}_{\mathrm{eff}}(H,d,\Delta)$  
differs from $H_{\mathrm{eff}}(H,d)$ only by an energy shift of
magnitude $\Delta$. The eigenstates of
$\widetilde{H}_{\mathrm{eff}}(H,d,\Delta)$ are identical to the
eigenstates of $H_{\mathrm{eff}}(H,d)$, as are all the gaps between
eigenenergies. The rest of this paper is devoted to showing that,
for any $k$-local Hamiltonian $H^{\mathrm{comp}}$ acting on $n$
qubits, there exists some function $f(\lambda)$ such that 
\begin{equation}
\label{result}
\widetilde{H}_{\mathrm{eff}}(H^{\mathrm{gad}}_+,2^n,f(\lambda))
= \frac{-k (-\lambda)^k}{(k-1)!}
H^{\mathrm{comp}} \otimes P_+ + \mathcal{O}(\lambda^{k+1})
\end{equation}
for sufficiently small $\lambda$. Here $P_+$ is an operator acting on
the ancilla registers, projecting each one into the state
$\ket{+}$. To obtain equation \ref{result} we use a formulation of
degenerate perturbation theory due to Bloch~\cite{Bloch, Messiah},
which we describe in the next section.

\section{Perturbation Theory}
\label{perturbation}

Suppose we have a Hamiltonian of the form 
\begin{equation}
H = H^{(0)} + \lambda V,
\end{equation}
where $H^{(0)}$ has a $d$-dimensional degenerate ground space
$\mathcal{E}^{(0)}$ of energy zero. As discussed in~\cite{Kato, Messiah},
the effective Hamiltonian for the $d$ lowest eigenstates of $H$ can be
obtained directly as a perturbation series in $V$. However, for our
purposes it is more convenient to use an indirect method due to
Bloch~\cite{Bloch, Messiah}, which we now describe. As shown in
appendix \ref{convergence}, the perturbative expansions converge
provided that
\begin{equation}
\label{conv_cond}
\| \lambda V \| < \frac{\gamma}{4},
\end{equation}
where $\gamma$ is the energy gap between the eigenspace in question
and the next nearest eigenspace, and $\| \cdot \|$ denotes the
operator norm\footnote{For any linear operator $M$,
\[
\| M \| \equiv \max_{|\braket{\psi}{\psi}| = 1} | \bra{\psi} M
\ket{\psi} |.
\]}.

Let $\ket{\psi_1}, \ldots, \ket{\psi_d}$ be the $d$ lowest energy
eigenstates of $H$, and let $E_1,\ldots,E_d$
be their energies. For small perturbations,
$\ket{\psi_1},\ldots,\ket{\psi_d}$ lie primarily within
$\mathcal{E}^{(0)}$. Let
\begin{equation}
\ket{\alpha_j} = P_0 \ket{\psi_j},
\end{equation}
where $P_0$ is the projector onto $\mathcal{E}^{(0)}$. For $\lambda$
satisfying \ref{conv_cond}, the  vectors $\ket{\alpha_1}, \ldots,
\ket{\alpha_d}$ are linearly independent, and there exists
a linear operator $\cu$ such that 
\begin{equation}
\cu \ket{\alpha_j} = \ket{\psi_k} \quad \textrm{for $j=1,2,\ldots, d$}
\end{equation}
and
\begin{equation}
\cu \ket{\phi} = 0 \quad \mathrm{for} \quad \ket{\phi} \in
\mathcal{E}^{(0)\perp}.
\end{equation}
Similarly, let $\cu^{-1}$ be the operator satisfying
\begin{equation}
\cu^{-1} \ket{\psi_j} = \ket{\alpha_j} \quad \textrm{for
  $j=1,2,\ldots, d$}
\end{equation}
and,
\begin{equation}
\cu^{-1} \ket{\phi} = 0 \quad \mathrm{for } \quad \ket{\phi} \in
\mathcal{E}^\perp.
\end{equation}
(Here $\mathcal{E}^\perp$ is defined implicitly by equation \ref{E-quation}.)
Note that $\cu$ is not invertible on the entire Hilbert space, as it
has a large kernel. However, if we consider $\cu$ as a linear
transformation from $\mathcal{E}^{(0)}$ to $\mathcal{E}$ then 
$\cu^{-1}:\mathcal{E} \to \mathcal{E}^{(0)}$ is its inverse. Also note
that, in general $\ket{\alpha_1},\ldots,\ket{\alpha_d}$ are neither
orthogonal nor normalized. Let
\begin{equation}
\label{afromu}
\ca = \lambda P_0 V \cu.
\end{equation}
As shown in~\cite{Messiah, Bloch} and recounted in appendix
\ref{Blochapp}, the right eigenvectors of $\ca$ are
$\ket{\alpha_1},\ldots,\ket{\alpha_d}$, and the corresponding
eigenvalues are $E_1, \ldots, E_d$. Thus $\ca$ encodes all information
about the low lying eigenstates of the perturbed Hamiltonian. More
precisely,
\begin{equation}
\label{heff}
H_{\mathrm{eff}} = \cu \ca \cu^{-1}.
\end{equation}

$\ca$ has the following perturbative expansion. Let $S^l$
be the operator
\begin{equation}
\label{S}
S^l = \left\{ \begin{array}{ll}
\displaystyle \sum_{j \neq 0} \frac{P_j}{(-E^{(0)}_j)^l} & 
\textrm{if $l >  0$} \\ \\
\displaystyle -P_0 & \textrm{if $l = 0$} 
\end{array} \right.
\end{equation}
where $P_j$ is the projector onto the eigenspace of $H^{(0)}$ with
energy $E^{(0)}_j$. (Recall that $E^{(0)}_0 = 0$.) Then
\begin{equation}
\label{ca}
\ca = \sum_{m=1}^\infty \ca^{(m)},
\end{equation}
where
\begin{equation}
\label{caterms}
\ca^{(m)} = \lambda^m \sum_{(m-1)} P_0 V S^{l_1} V S^{l_2} \ldots V
S^{l_{m-1}} V P_0, 
\end{equation}
and the sum is over all nonnegative integers $l_1 \ldots l_{m-1}$
satisfying
\begin{eqnarray}
l_1 + \ldots + l_{m-1} & = & m-1 \\
l_1 + \ldots + l_p & \geq & p \quad (p = 1,2,\ldots,m-2).
\end{eqnarray}
For our purposes, we do not need the perturbative expansions for $\cu$
and $\cu^{-1}$ except to note that
\begin{eqnarray}
\cu & = & P_0 + \mathcal{O}(\lambda) \\
\cu^{-1} & = & P_0 + \mathcal{O}(\lambda).
\end{eqnarray}
For completeness, we provide derivations for the expansions of $\cu$
and $\ca$ in appendix \ref{Blochapp}. In appendix \ref{convergence} we
prove that condition \ref{conv_cond} suffices to ensure
convergence. The advantage of the method of~\cite{Bloch} over the
direct approach of~\cite{Kato} is that $\ca$ is an operator whose
support is strictly within $\mathcal{E}^{(0)}$, which makes some of
the calculations more convenient. 

\section{Analysis of the Gadget Hamiltonian}
\label{section_analysis}

Before analyzing $H^{\mathrm{gad}}$ for a general $k$-local
Hamiltonian, we first consider the case where $H^{\mathrm{comp}}$
has one coefficient $c_s = 1$ and all the rest equal to zero. That
is,
\begin{equation}
H^{\mathrm{comp}} = \sigma_1 \sigma_2 \ldots \sigma_k, 
\end{equation}
where for each $j$, $\sigma_j = \hat{n}_j \cdot \vec{\sigma}_j$ for some
unit vector $\hat{n}_j$ in $\mathbb{R}^3$. The corresponding gadget
Hamiltonian is thus 
\begin{equation}
H^{\mathrm{gad}} = H^{\mathrm{anc}} + \lambda V, 
\end{equation}
where
\begin{equation}
\label{nonpert_recap}
H^{\mathrm{anc}} = \sum_{1 \leq i < j \leq k} \frac{1}{2}(I-Z_i Z_j),
\end{equation}
and
\begin{equation}
\label{pert_recap}
V = \sum_{j=1}^k \sigma_j \otimes X_j.
\end{equation}
Here $\sigma_j$ acts on the $j\th$ computational qubit, and $X_j$ and
$Z_j$ are the Pauli $X$ and $Z$ operators acting on the $j\th$ ancilla
qubit. We use $k\th$ order perturbation theory to show that
$\widetilde{H}^{\mathrm{eff}}(H^{\mathrm{gad}}_+,2^k,\Delta)$
approximates $H^{\mathrm{comp}}$ for appropriate $\Delta$.

We start by calculating $\ca$ for $H_+^{\mathrm{gad}}$. For
$H^{\mathrm{anc}}$, the energy gap is 
$\gamma = k-1$, and $\| V \| = k$, so by condition \ref{conv_cond}, we
can use perturbation theory provided $\lambda$ satisfies 
\begin{equation}
\lambda < \frac{k-1}{4k}.
\end{equation}
Because all terms in $\ca$ are sandwiched by $P_0$ operators, the
nonzero terms in $\ca$ are ones in which the $m$ powers of $V$ take a
state in $\mathcal{E}^{(0)}$ and return it to
$\mathcal{E}^{(0)}$. Because we are working in the $+1$ eigenspace of
$X^{\otimes k}$, an examination of equation \ref{nonpert_recap} shows that
$\mathcal{E}^{(0)}$ is the span of the states in which the ancilla qubits
are in the state $\ket{+}$. Thus, $P_0 = I \otimes P_+$, where $P_+$
acts only on the ancilla qubits, projecting them onto the state
$\ket{+}$. Each term in $V$ flips one ancilla qubit. To return to
$\mathcal{E}^{(0)}$, the powers of $V$ must either flip some ancilla
qubits and then flip them back, or they must flip all of them. The
latter process occurs at $k\th$ order and gives rise
to a term that mimics $H^{\mathrm{comp}}$. The former process occurs
at many orders, but at orders $k$ and lower gives rise only to terms
proportional to $P_0$.

As an example, let's examine $\ca$ up to second order for $k > 2$.
\begin{equation}
\ca^{(\leq 2)} = \lambda P_0 V P_0 + \lambda^2 P_0 V S^1 V P_0
\end{equation}
The term $P_0 V P_0$ is zero, because $V$ kicks the state out of
$\mathcal{E}^{(0)}$. By equation \ref{pert_recap} we see that applying
$V$ to a state in the ground space yields a state in the energy $k-1$
eigenspace. Substituting this denominator into $S^1$ yields
\begin{equation}
\ca^{(2)} = -\frac{\lambda^2}{k-1} P_0 V^2 P_0.
\end{equation} 
Because $V$ is a sum, $V^2$ consists of the squares of individual
terms of $V$ and cross terms. The cross terms flip two ancilla qubits,
and thus do not return the state to the ground space. The
squares of individual terms are proportional to the identity, thus
\begin{equation}
\ca^{(2)} = \lambda^2 \alpha_2 P_0
\end{equation}
for some $\lambda$-independent constant $\alpha_2$. Similarly, at any
order $m < k$, the only terms in $V^m$ which project back to
$\mathcal{E}^{(0)}$ are those arising from squares of individual
terms, which are proportional to the identity. Thus, up to order
$k-1$,
\begin{equation}
\ca^{(\leq k-1)} = \left( \sum_m \alpha_m \lambda^m \right)
P_0
\end{equation}
where the sum is over even $m$ between zero and $k-1$ and
$\alpha_0, \alpha_2,\ldots$ are the corresponding coefficients.

At $k\th$ order there arises another type of term. In $V^k$ there are
$k$-fold cross terms in which each of the terms in $V$ appears
once. For example, there is the term
\begin{equation}
\label{goodterm}
\lambda^k P_0 (\sigma_1 \otimes X_1)S^1 (\sigma_2
\otimes X_2) S^1 \ldots S^1 (\sigma_k \otimes
X_k) P_0
\end{equation}
The product of the energy denominators occurring in the $S^1$ operators
is
\begin{equation}
\prod_{j=1}^{k-1} \frac{1}{-j (k-j)} = \frac{(-1)^{k-1}}{((k-1)!)^2}.
\end{equation}
Thus, this term is
\begin{equation}
\frac{(-1)^{k-1} \lambda^k}{((k-1)!)^2} P_0 (\sigma_1 \otimes X_1) (\sigma_2
\otimes X_2) \ldots (\sigma_k \otimes X_k) P_0,
\end{equation}
which can be rewritten as
\begin{equation}
\frac{-(-\lambda)^k}{((k-1)!)^2} P_0 (\sigma_1 \sigma_2 \ldots \sigma_k
\otimes X^{\otimes k}) P_0.
\end{equation}
This term mimics $H^{\mathrm{comp}}$. The fact that all the $S$
operators in this term are $S^1$ is a general feature. Any term in
$\ca^{(k)}$ where $l_1 \ldots l_{k-1}$ are not all equal to 1
either vanishes or is proportional to $P_0$. This is because such 
terms contain $P_0$ operators separated by fewer than $k$ powers of
$V$, and thus the same arguments used for $m < k$ apply.

There are a total of $k!$ terms of the type shown in expression
\ref{goodterm}. 
Thus, up to $k\th$ order
\begin{equation}
\ca^{(\leq k)} = f(\lambda) P_0 + \frac{-k (-\lambda)^k}{(k-1)!} P_0
(\sigma_1 \sigma_2 \ldots \sigma_k \otimes X^{\otimes k}) P_0,
\end{equation}
which can be written as
\begin{equation}
\label{mpauli}
\ca^{(\leq k)} = f(\lambda) P_0 + \frac{-k (-\lambda)^k}{(k-1)!} P_0
(H^{\mathrm{comp}} \otimes X^{\otimes k}) P_0
\end{equation}
where $f(\lambda)$ is some polynomial in $\lambda$. By equations
\ref{mpauli} and \ref{heff},
\begin{eqnarray}
H_{\mathrm{eff}}(H_+^{\mathrm{gad}},2^k) & = & \cu f(\lambda) P_0
                       \cu^{-1} + \cu \left[ \frac{-k (-\lambda)^k}{(k-1)!} P_0
                       (H^{\mathrm{comp}} \otimes X^{\otimes k}) P_0 
                       +\mathcal{O}(\lambda^{k+1}) \right] \cu^{-1}
                       \nonumber \\ 
                       & = & f(\lambda) \Pi + \cu 
                       \left[\frac{-k (-\lambda)^k}{(k-1)!} 
                       P_0 (H^{\mathrm{comp}} \otimes X^{\otimes k}) P_0
                       + \mathcal{O}(\lambda^{k+1}) \right] \cu^{-1}
\end{eqnarray}
since $\cu P_0 \cu^{-1} = \Pi$. Thus,
\begin{equation}
\label{halmost}
\widetilde{H}_{\mathrm{eff}}(H_+^{\mathrm{gad}},2^k,f(\lambda)) = \cu \left[
                       \frac{-k (-\lambda)^k}{(k-1)!}
                       P_0(H^{\mathrm{comp}} \otimes X^{\otimes k})P_0
                       +\mathcal{O}(\lambda^{k+1}) \right] \cu^{-1}.
\end{equation}
In equation \ref{halmost}, we can approximate $\cu$ and $\cu^{-1}$ as
$P_0$ since the higher order corrections to $\cu$ give rise to terms
of order  $\lambda^{k+1}$ and higher in the expression for
$\widetilde{H}_{\mathrm{eff}}(H_+^{\mathrm{gad}},2^k,f(\lambda))$. Thus,
\begin{equation}
\widetilde{H}_{\mathrm{eff}}(H_+^{\mathrm{gad}},2^k,f(\lambda)) =
                        \frac{-k (-\lambda)^k}{(k-1)!} P_0 
                        (H^{\mathrm{comp}} \otimes X^{\otimes k}) P_0 +
                        \mathcal{O}(\lambda^{k+1}).
\end{equation}
Using $P_0 = I \otimes P_+$ we rewrite this as
\begin{equation}
\widetilde{H}_{\mathrm{eff}}(H_+^{\mathrm{gad}},2^k,f(\lambda)) = 
                        \frac{-k (-\lambda)^k}{(k-1)!}
                        H^{\mathrm{comp}} \otimes P_+ +
                        \mathcal{O}(\lambda^{k+1}).
\end{equation}

Now let's return to the general case where $H^{\mathrm{comp}}$ is a
linear combination of $k$-local terms with arbitrary coefficients
$c_s$, as described in equation \ref{hcomp}. Now that we have
gadgets to obtain $k$-local effective interactions, it is tempting to
eliminate one $k$-local interaction at a time, by introducing
corresponding gadgets one by one. However, this approach does not lend
itself to simple analysis by degenerate perturbation theory. This is
because the different $k$-local terms in general act on
overlapping sets of qubits. Hence, we instead consider 
\begin{equation}
\label{vgad}
V^{\mathrm{gad}} = \sum_{s=1}^r V_s
\end{equation}
as a single perturbation, and work out the effective Hamiltonian in
powers of this operator. The unperturbed part of the total gadget
Hamiltonian is thus
\begin{equation}
H^{\mathrm{anc}} = \sum_{s=1}^r H_s^{\mathrm{anc}},
\end{equation}
which has energy gap $\gamma = k-1$. The full Hamiltonian is
\begin{equation}
H^{\mathrm{gad}} = H^{\mathrm{anc}} + \lambda V^{\mathrm{gad}},
\end{equation}
so the perturbation series is guaranteed to converge under the
condition
\begin{equation}
\label{lambda_size}
\lambda < \frac{k-1}{4 \| V^{\mathrm{gad}} \|}
\end{equation}
As mentioned previously, we will work only within the simultaneous
$+1$ eigenspace of the $X^{\otimes k}$ operators acting on each of the
ancilla registers. In this subspace, $H^{\mathrm{anc}}$ has degeneracy
$2^n$ which gets split by the perturbation $\lambda V$ so that it
mimics the spectrum of $H^{\mathrm{comp}}$.

Each $V_s$ term couples to a different ancilla register. Hence, any
cross term between different $V_s$ terms flips some ancilla
qubits in one register and some ancilla qubits in another. Thus, at
$k\th$ order, non-identity cross terms between different $s$ cannot
flip all $k$ ancilla qubits in any given ancilla register, and they
are thus projected away by the $P_0$ operators appearing in the
formula for $\ca$. Hence the perturbative analysis proceeds just as it
did when there was only a single nonzero $c_s$, and one finds,
\begin{equation}
\widetilde{H}_{\mathrm{eff}}(H_+^{\mathrm{gad}},2^n,f(\lambda)) =
                        \frac{-k (-\lambda)^k}{(k-1)!}
                        P_0 \left( \sum_{s=1}^r c_s H_s 
                        \otimes X_s^{\otimes k} \right) P_0 +
                        \mathcal{O}(\lambda^{k+1}), 
\end{equation}
where $X_s^{\otimes k}$ is the operator $X^{\otimes k}$ acting
on the register of $k$ ancilla qubits corresponding to a given
$s$, and $f(\lambda)$ is some polynomial in $\lambda$ of degree at
most $k$. Note that coefficients in the polynomial $f(\lambda)$ depend
on $H^{\mathrm{comp}}$. As before, this can be rewritten as
\begin{equation}
\label{finalheff}
\widetilde{H}_{\mathrm{eff}}(H_+^{\mathrm{gad}},2^n,f(\lambda)) =  
                \frac{-k (-\lambda)^k}{(k-1)!} H^{\mathrm{comp}}
                \otimes P_+ + \mathcal{O}(\lambda^{k+1}),
\end{equation}
where $P_+$ acts only on the ancilla registers, projecting them
all into the $\ket{+}$ state. Hence, as asserted in section
\ref{intro}, the 2-local gadget Hamiltonian $H^{\mathrm{gad}}$
generates effective interactions which mimic the $k$-local Hamiltonian
$H^{\mathrm{comp}}$. 

For a polynomial time adiabatic quantum computation one needs a
Hamiltonian that varies smoothly in time and has an eigenvalue gap at
worst polynomially small. Let $H(t)$ be a $k$-local Hamiltonian of
this type. For each time $t$ one can construct the corresponding
instantaneous gadget Hamiltonian $H^{\mathrm{gad}}(t)$ as described in
section \ref{intro} and equation \ref{lambda_size}. It is not hard to
show that $H^{\mathrm{gad}}(t)$ varies smoothly in time and has a gap
that is polynomial in $n$ for any fixed $k$. Thus
$H^{\mathrm{gad}}(t)$ is a 2-local polynomial-time adiabatic algorithm
that simulates the original $k$-local algorithm $H(t)$. In addition to
adiabatic quantum computation we expect that $k\th$ order gadgets may
have many other applications in quantum computation, such as proving
QMA-completeness.

\section{Numerical Examples}

\begin{figure}[!htbp]
\begin{center}
\includegraphics[width=0.65\textwidth]{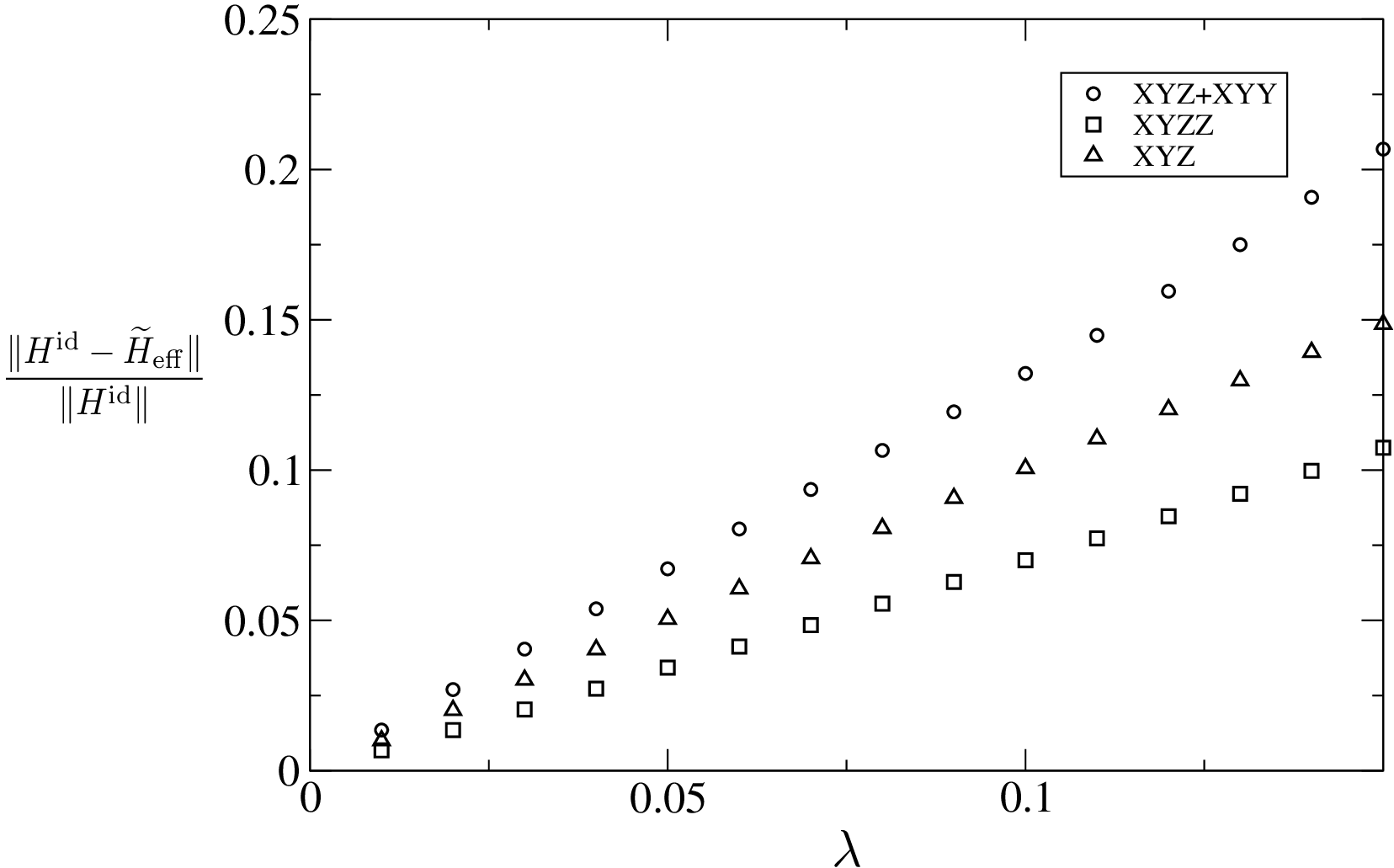}
\caption{\label{ratio} Here the ratio of the error terms to the ideal
  Hamiltonian $H^{\mathrm{id}} \equiv \frac{-k (-\lambda)^k}{(k-1)!}
  H^{\mathrm{comp}}$ is plotted. We examine three examples, a third
  order gadget simulating a single $XYZ$ interaction, a third order
  gadget simulating a pair of interactions $XYZ+XYY$, and a fourth
  order gadget simulating a fourth order interaction
  $XYZZ$. Here $\widetilde{H}_{\mathrm{eff}}$ is calculated by direct
  numerical computation without using perturbation theory. As expected
  the ratio of the norm of the error terms to $H^{\mathrm{id}}$ goes
  linearly to zero with shrinking $\lambda$.}
\end{center}
\end{figure}

In this section we numerically examine the performance of
perturbative gadgets in some small examples. As shown in section
\ref{section_analysis}, the shifted effective Hamiltonian is that given in
equation \ref{finalheff}. We define
\begin{equation}
H^{\mathrm{id}} \equiv \frac{-k (-\lambda)^k}{(k-1)!}
H^{\mathrm{comp}} \otimes P_+.
\end{equation}
$\widetilde{H}_{\mathrm{eff}}$ consists of the ideal piece
$H^{\mathrm{id}}$, which is of order $\lambda^k$, plus an error term of
order $\lambda^{k+1}$ and higher. For sufficiently small $\lambda$ these error
terms are therefore small compared to the $H^{\mathrm{id}}$ term which
simulates $H^{\mathrm{comp}}$. Indeed, by a calculation very similar
to that which appears in appendix \ref{convergence}, one can easily
place an upper bound on the norm of the error terms. However, in
practice the actual size of the error terms may be smaller than this
bound. To examine the error magnitude in practice, we plot
$ \frac{\| H^{\mathrm{id}} - \widetilde{H}_{\mathrm{eff}}
  \|}{\|H^{\mathrm{id}}\|}$ in figure \ref{ratio} using direct 
numerical computation of $\widetilde{H}_{\mathrm{eff}}$ without
perturbation theory. $f(\lambda)$ was calculated analytically for
these examples. In all cases the ratio of $\| H^{\mathrm{id}} -
  \widetilde{H}_{\mathrm{eff}}\|$ to $\|H^{\mathrm{id}}\|$ scales
approximately linearly with $\lambda$, as one expects since the
error terms are of  order $\lambda^{k+1}$ and higher, whereas
$H^{\mathrm{id}}$ is of order $\lambda^k$.

\section{Acknowledgements} 
We thank Ognyan Oreshkov, Sergey Bravyi, Jake Taylor, Dave Bacon,
Michael Levin, Barbara Terhal, Daniel Lidar, Mark Rudner, Jacob
Biamonte, and Peter Love for useful discussions. SJ thanks the Army
Research Office (ARO) and Disruptive Technology Office (DTO) for
support under their QuaCGR program. EF gratefully acknowledges support
from ARO contract W911NF-04-0216, and the Keck foundation.

\appendix

\section{Derivation of Perturbative Formulas}
\label{Blochapp}

In this appendix we give a self-contained presentation of the
derivations for the method of degenerate perturbation theory used in
this paper. We closely follow Bloch~\cite{Bloch}. Given a Hamiltonian
of the form
\begin{equation}
H = H^{(0)} + \lambda V
\end{equation}
we wish to find the effective Hamiltonian induced by the perturbation
$\lambda V$ on the ground space of $H^{(0)}$. In what follows, we
assume that the ground space of $H^{(0)}$ has energy zero. This
simplifies notation, and the generalization to nonzero ground energy
is straightforward. To further simplify notation we define
\begin{equation}
\hv = \lambda V.
\end{equation}

Suppose the ground space of $H^{(0)}$ is $d$-dimensional and denote it
by $\mathcal{E}^{(0)}$. Let
$\ket{\psi_1}, \ldots, \ket{\psi_d}$ be the perturbed eigenstates
arising from the splitting of this degenerate ground
space, and let $E_1, \ldots, E_d$ be their energies. Furthermore,
let $\ket{\alpha_j} = P_0 \ket{\psi_j}$ where $P_0$ is the projector
onto the unperturbed ground space of $H^{(0)}$. If $\lambda$ is
sufficiently small, $\ket{\alpha_1},\ldots,\ket{\alpha_d}$ are
linearly independent, and we can define an operator $\cu$ such that
\begin{equation}
\label{cudef}
\cu \ket{\alpha_j} = \ket{\psi_j}
\end{equation}
and
\begin{equation}
\label{uproj}
\cu \ket{\phi} = 0 \quad \forall \ket{\phi} \in \mathcal{E}^{(0)\perp}.
\end{equation}

Now let $\ca$ be the operator
\begin{equation}
\label{pvcu}
\ca = P_0 \hv \cu.
\end{equation}
$\ca$ has $\ket{\alpha_1}, \ldots, \ket{\alpha_d}$ as right
eigenvectors, and $E_1, \ldots, E_d$ as corresponding eigenvalues. To
see this, note that since $H^{(0)}$ has zero ground state energy
\begin{equation}
\label{zerotrick}
P_0 \hv = P_0 (H^{(0)}+\hv) = P_0 H.
\end{equation}
Thus,
\begin{eqnarray}
\ca \ket{\alpha_j} & = & P_0 \hv \cu \ket{\alpha_j} \nonumber \\
& = & P_0 \hv \ket{\psi_j} \nonumber \\
& = & P_0 H \ket{\psi_j} \nonumber \\
& = & P_0 E_j \ket{\psi_j} \nonumber \\
& = & E_j \ket{\alpha_j}.
\end{eqnarray}
The essential task in this formulation of
degenerate perturbation theory is to find a perturbative expansion for
$\cu$. From $\cu$ one can obtain $\ca$ by equation \ref{pvcu}. Given
$\ca$, one can easily claulate its right eigenvectors
$\ket{\alpha_1},\ldots,\ket{\alpha_d}$ and the corresponding
eigenvalues $E_1,\ldots, E_d$. Then, by applying $\cu$ to
$\ket{\alpha_j}$ one obtains $\ket{\psi_j}$. So, given a
perturbative formula for $\cu$, all quantities of interest can be
calculated. 

In the remainder of this appendix we will derive the following
\begin{equation}
\label{cu}
\cu = P_0 + \sum_{m=1}^\infty \cu^{(m)},
\end{equation}
where
\begin{equation}
\label{cuterms}
\cu^{(m)} = \lambda^m \sum_{(m)} S^{l_1} V S^{l_2} V \ldots V
S^{l_m} V P_0,
\end{equation}
$S^l$ is as given in equation \ref{S}, and the sum is over all sets of
nonnegative integers $l_1,\ldots,l_m$ such that
\begin{eqnarray}
l_1 + \ldots + l_m & = & m \\
l_1 + \ldots + l_p & \geq & p \quad (p=1,2,\ldots,m-1).
\end{eqnarray}
To derive this, we start with Schr\"odinger's equation:
\begin{equation}
\label{schro}
H \ket{\psi_j} = E_j \ket{\psi_j}.
\end{equation}
By equation \ref{zerotrick}, left-multiplying this by $P_0$ yields
\begin{equation}
\label{pnaught}
P_0 \hv\ket{\psi_j} = E_j \ket{\alpha_j}.
\end{equation}
By equation \ref{uproj},
\begin{equation}
\label{icup}
\cu P_0 = \cu.
\end{equation}
Thus left-multiplying equation \ref{pnaught} by $\cu$ yields
\begin{equation}
\label{cuv}
\cu \hv \ket{\psi_j} = E_j \ket{\psi_j}.
\end{equation}
By subtracting \ref{cuv} from \ref{schro} we obtain
\begin{equation}
(H - \cu \hv) \ket{\psi_j} = 0.
\end{equation}
The span of $\ket{\psi_j}$ we call $\mathcal{E}$. For any state
$\ket{\beta}$ in $\mathcal{E}$ we have
\begin{equation}
(H - \cu \hv) \ket{\beta} = 0.
\end{equation}
Since $\cu \ket{\gamma} \in \mathcal{E}$ for any state $\ket{\gamma}$,
it follows that
\begin{equation}
(H - \cu \hv) \cu = 0.
\end{equation}
This equation can be rewritten as
\begin{equation}
\label{ho}
H^{(0)} \cu = -\hv \cu + \cu \hv \cu.
\end{equation}
Defining $Q_0 = \id - P_0$ we have
\begin{equation}
\label{icuq}
\cu = P_0 \cu + Q_0 \cu.
\end{equation}
Substituting this into the left side of \ref{ho} yields
\begin{equation}
H^{(0)} Q_0 \cu = - \hv \cu + \cu \hv \cu,
\end{equation}
because $H^{(0)} P_0 = 0$. In $\mathcal{E}^{(0)\perp}$, $H^{(0)}$ has a well
defined inverse and one can write
\begin{equation}
Q_0 \cu = -\frac{1}{H^{(0)}} Q_0 (\hv \cu - \cu \hv \cu). 
\end{equation}
Using equation \ref{icuq}, one obtains
\begin{equation}
\cu = P_0 \cu - \frac{1}{H^{(0)}} Q_0 ( \hv \cu - \cu \hv \cu ).
\end{equation}
By the definition of $\cu$ it is apparent that $P_0 \cu = P_0$, thus
this equation simplifies to
\begin{equation}
\label{imp}
\cu = P_0 - \frac{1}{H^{(0)}} Q_0 ( \hv \cu - \cu \hv \cu ).
\end{equation}

We now expand $\cu$ in powers of $\lambda$ (equivalently, in powers
of $\hv$), and denote the $m\th$ order term by
$\cu^{(m)}$. Substituting this expansion into equation 
\ref{imp} and equating terms at each order yields the following
recurrence relations.
\begin{eqnarray}
\label{rec1}
\cu^{(0)} & = & P_0 \\
\label{rec2}
\cu^{(m)} & = & -\frac{1}{H^{(0)}} Q_0 \left[ \hv \cu^{(m-1)} -
  \sum_{p=1}^{m-1} \cu^{(p)} \hv \cu^{(m-p-1)} \right] \quad (m=1,2,3\ldots)
\end{eqnarray}
Note that the sum over $p$ starts at $p=1$, not $p=0$. This is because
\begin{equation}
\frac{1}{H^{(0)}} Q_0 \cu^{(0)} = \frac{1}{H^{(0)}} Q_0 P_0 = 0.
\end{equation}
Let 
\begin{equation}
S^l = \left\{ \begin{array}{ll}
\frac{1}{\left(-H^{(0)}\right)^l} Q_0 & \textrm{if $l > 0$} \\
- P_0 & \textrm{if $l = 0$}
\end{array} \right. .
\end{equation}
$\cu^{(m)}$ is of the form
\begin{equation}
\cu^{(m)} = {\sum}' S^{l_1} \hv S^{l_2} \hv \ldots S^{l_m} \hv P_0,
\end{equation}
where $\sum'$ is a sum over some subset of $m$-tuples $(l_1, l_2,
\ldots, l_m)$ such that
\begin{equation}
l_i \geq 0 \quad (i=1,2,\ldots, m)
\end{equation}
\begin{equation}
l_1 + l_2 + \ldots + l_m = m.
\end{equation}
The proof is an easy induction. $\cu^{(0)}$ clearly satisfies this,
and we can see that if $\cu^{(j)}$ has these properties for all
$j < m$, then by recurrence \ref{rec2}, $\cu^{(m)}$ also has
these properties. 

All that remains is to prove that the subset of allowed $m$-tuples
appearing in the sum $\sum'$ are exactly those which satisfy
\begin{equation}
\label{convex}
l_1 + \ldots + l_p \geq p \quad (p=1,2,\ldots,m-1).
\end{equation}

\begin{figure}
\begin{center}
\includegraphics[width=0.22\textwidth]{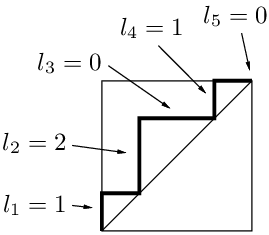}
\caption{\label{stairs} From a given $m$-tuple $(l_1,l_2,\ldots, l_m)$
  we construct a corresponding stairstep diagram by making the $j\th$
  step have height $l_j$, as illustrated above.}
\end{center}
\end{figure}

Following~\cite{Bloch}, we do this by introducing stairstep
diagrams to represent the $m$-tuples, as shown in figure \ref{stairs}.
The $m$-tuples with property \ref{convex} correspond to diagrams in
which the steps lie above the diagonal. Following~\cite{Bloch} we call
these convex diagrams. Thus our task is to prove that the sum $\sum'$
is over all and only the convex diagrams. To do this, we consider
the ways in which convex diagrams of order $m$ can be constructed from
convex diagrams of lower order. We then relate this to the way
$\cu^{(m)}$ is obtained from lower order terms in the recurrence
\ref{rec2}.

\begin{figure}
\begin{center}
\includegraphics[width=0.5\textwidth]{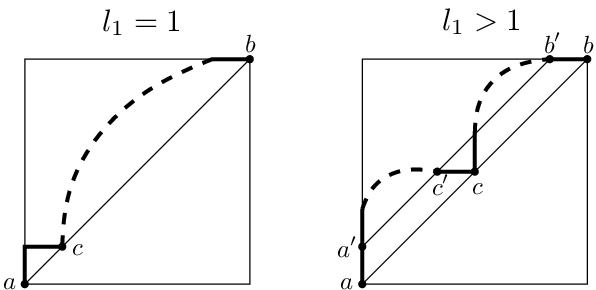}
\caption{\label{cases} A convex diagram must have either $l_1 = 1$ or
  $l_1 > 1$. In either case, the diagram can be decomposed as a
  concatenation of lower order convex diagrams.}
\end{center}
\end{figure}

In any convex diagram, $l_1 \geq 1$. We now consider the two cases
$l_1=1$ and $l_1 > 1$. In the case that $l_1 = 1$, the diagram is as
shown on the left in figure \ref{cases}.
In any convex diagram of order $m$ with $l_1 = 1$, there is an
intersection with the diagonal after one step, at the point
that we have labelled $c$. The diagram from $c$ to $b$ is a convex
diagram of order $m-1$. Conversely, given any convex diagram of order
$m-1$ we can construct a convex diagram of order $m$ by adding one
step to the beginning. Thus the convex diagrams of order $m$ with $l_1
= 1$ correspond bijectively to the convex diagrams of order $m-1$.

The case $l_1 > 1$ is shown in figure \ref{cases} on the right. Here
we introduce the line from $a'$ to $b'$, which is parallel to the
diagonal, but higher by one step. Since the diagram must end at $b$,
it must cross back under $a'b'$ at some point. We'll label the first
point at which it does so as $c'$. In general, $c'$ can equal
$b'$. The curve going from $a'$ to $c'$ is a convex diagram of order
$p$ with $1 \leq p \leq m-1$, and the curve going from $c$ to $b$ is a
convex diagram of order $n-p-1$ (which may be order zero if $c'$ =
$b'$). Since $c'$ exists and is unique, this establishes a bijection
between the convex diagrams of order $m$ with $l_1 > 1$, and the set
of the pairs of convex diagrams of orders $p$ and $n-p-1$, for $1 \leq
p \leq n-1$.

Examining the recurrence \ref{rec2}, we see that the $l_1 = 1$
diagrams are exactly those which arise from the term
\begin{equation}
\frac{Q_0}{H^{(0)}} \hv \cu^{(m-1)}
\end{equation}
and the $l_1 > 1$ diagrams are exactly those which arise from the term
\begin{equation}
\frac{Q_0}{H^{(0)}} \sum_{p=1}^{m-1} \cu^{(p)} \hv \cu^{(n-p-1)}.
\end{equation}
which completes the proof that $\sum'$ is over the $m$-tuples
satisfying equation \ref{convex}.

\section{Convergence of Perturbation Series}
\label{convergence}

Here we show that the perturbative expansion for $\cu$
given in equation \ref{cu} converges for
\begin{equation}
\| \lambda V \| < \frac{\gamma}{4}.
\end{equation}
By equation \ref{afromu}, the convergence of $\cu$ also implies the
convergence of $\ca$. Applying the triangle inequality to equation
\ref{cu} yields
\begin{equation}
\label{tri1}
\| \cu \| \leq 1 + \sum_{m=1}^{\infty} \| \cu^{(m)} \|.
\end{equation}
Substituting in equation \ref{cuterms} and applying the triangle
inequality again yields
\begin{equation}
\| \cu \| \leq 1 + \sum_{m=1}^{\infty} \lambda^m \sum_{(m)} \| S^{l_1}
\ldots V S^{l_m} V P_0 \|.
\end{equation}
By the submultiplicative property of the operator norm,
\begin{equation}
\label{intermed}
\| \cu \| \leq 1 + \sum_{m=1}^{\infty} \lambda^m \sum_{(m)} \| S^{l_1}
\| \cdot \| V \| \ldots \| V \| \cdot \| S^{l_m} \| \cdot \| V \|
\cdot \| P_0 \|.
\end{equation}
$\| P_0 \| = 1$, and by equation \ref{S} we have
\begin{equation}
\| S^l \| = \frac{1}{(E_1^{(0)})^l} = \frac{1}{\gamma^l}.
\end{equation}
Since the sum in equation \ref{intermed} is over $l_1 + \ldots +
l_m = m$, we have
\begin{equation}
\| \cu \| \leq 1 + \sum_{m=1}^\infty \sum_{(m)} \frac{\| \lambda V
  \|^m}{\gamma^m}. 
\end{equation}
The sum $\sum_{(m)}$ is over a subset of the $m$-tuples
adding up to $m$. Thus, the number of terms in this sum is less than
the number of ways of obtaining $m$ as a sum of $m$ nonnegative
integers. By elementary combinatorics, the number of ways to obtain
$n$ as a sum of $r$ nonnegative integers is $\binom{n+r-1}{n}$, thus
\begin{equation}
\label{notgeom}
\| \cu \| \leq 1 + \sum_{m=1}^\infty \binom{2m-1}{m} \frac{\| \lambda
  V \|^m}{\gamma^m}.
\end{equation}
Since
\begin{equation}
\sum_{j=0}^{2m-3} \binom{2m-1}{j} = 2^{2m-1},
\end{equation}
we have
\begin{equation}
\binom{2m-1}{m} \leq 2^{2m-1}.
\end{equation}
Substituting this into equation \ref{notgeom} converts it into a
convenient geometric series:
\begin{equation}
\| \cu \| \leq 1 + \sum_{m=1}^\infty 2^{2m-1} \frac{\| \lambda V
  \|^m}{\gamma^m}.
\end{equation}
This series converges for
\begin{equation}
\frac{4 \| \lambda V \|}{\gamma} < 1.
\end{equation}

\bibliography{gadgets}

\end{document}